\documentclass[sn-mathphys]{sn-jnl}

\usepackage{setspace}
\usepackage{array}
\usepackage{graphicx}
\usepackage{subcaption}

\jyear{2021}%

\theoremstyle{thmstyleone}%
\newtheorem{theorem}{Theorem}
\newtheorem{lemma}[theorem]{Lemma}%

\theoremstyle{thmstyletwo}%
\newtheorem{example}{Example}%

\theoremstyle{thmstylethree}%

\begin{document}

\title[HPC acceleration of large $(\min,+)$ matrix products]{HPC acceleration of large $(\min,+)$ matrix products to compute domination-type parameters in graphs}

\author[1,3]{\fnm{E.M.} \sur{Garz\'on}}\email{gmartin@ual.es}

\author[1,3]{\fnm{J.A.} \sur{Mart\'inez}}\email{jmartine@ual.es}

\author[1,3]{\fnm{J.J.} \sur{Moreno}}\email{juanjomoreno@ual.es}

\author*[2,3]{\fnm{M.L.} \sur{Puertas}}\email{mpuertas@ual.es}
\equalcont{All the authors contributed equally to this work.}

\affil[1]{\orgdiv{Department of Computer Sciences}, \orgname{Universidad de Almería}, \country{Spain}}

\affil[2]{\orgdiv{Department of Mathematics}, \orgname{Universidad de Almería}, \country{Spain}}

\affil[3]{\orgdiv{Agrifood Campus of International Excellence (ceiA3)}, \orgname{Universidad de Almería}, \country{Spain}}

\abstract{
{\color{black}The computation of the domination-type parameters is a challenging problem in Cartesian product graphs. We present an algorithmic method to compute the $2$-domination number of the Cartesian product of a path with small order and any cycle, involving the $(\min,+)$ matrix product. We establish some theoretical results that provide the algorithms necessary to compute that parameter, and the main challenge to run such algorithms comes from the large size of the matrices used, which makes it necessary to improve the techniques to handle these objects. We analyze the performance of the algorithms on modern multicore CPUs and on GPUs and we show the advantages over the sequential implementation. The use of these platforms allows us to compute the $2$-domination number of cylinders such that their paths have at most $12$ vertices. }
}


\maketitle

\section{Introduction}\label{section:intro}

The $(\min,+)$ matrix algebra~\cite{Pin1998}, also called tropical algebra, replaces addition and multiplication with minimization and addition respectively. The use of this algebra is currently in expansion and it is involved in several disciplines of great interest, {\color{black}for instance finite automata~\cite{Pin1998}, statistics~\cite{Omanovic21}, phylogenetics~\cite{Speyer2009}, optimization of graph parameters~\cite{Klavzar1996}, integer programming~\cite{Butkovic2019}, and other optimization problems~\cite{Krivulin2015}.} However, the computational demands of such computations are unapproachable when the dimensions of the corresponding matrices are large. To overcome this drawback the modern multicore CPUs and GPUs can be exploited as High-Performance Computing (HPC) platforms to accelerate and widen the dimensions of such operations. In this work, the analysis of the domination-type parameters in graphs  is chosen as an interesting example where sequences of large $(\min, +)$ matrix  products are involved.

The use of graphs as a tool to model problems in networks has been widely studied. Among such problems, the efficient location of resources in a network can be approached by means of the domination-type parameters
in graphs. A dominating set in a graph $G$ is a vertex subset $S$ such that each vertex not in $S$ has at least one neighbor in it. The domination number of $G$, denoted by $\gamma(G)$, is the cardinal of a minimum
dominating set. We refer to~\cite{Haynes1998} for general information about these topics and, in particular, about their applications to network problems. Among the variations of this concept that can be
found in literature, we focus on the $2$-domination. A \emph{$2$-dominating set} is vertex subset $S\subseteq V(G)$ such that each vertex not in $S$ has at least two neighbors in it. The \emph{$2$-domination number}
$\gamma_2(G)$ is the minimum cardinal of a $2$-dominating set of $G$~\cite{Fink1985}. Some interesting applications of the $2$-domination in graphs such as the optimization of fault tolerant sensor networks, the facility location
problem and the data collection problem can be found in~\cite{Butjas2018}. {\color{black}Given a graph $G$ and a positive integer $k\leq \vert V(G)\vert $, the decision problem ``Is there a dominating set of G with at most k vertices?'' is NP-complete~\cite{Garey1979}}, even in bipartite and chordal graphs. However, it has been shown to be polynomial in trees and interval
graphs~\cite{Haynes1998}. {\color{black} In a similar way, the $2$-domination decision problem is to decide whether $G$ has a $2$-dominating set of cardinal at most $k\leq \vert V(G)\vert $. It is known that it is an NP-complete problem~\cite{Jacobson1989},} again even in bipartite and chordal graphs~\cite{Bean1994}.
Moreover, linear-time algorithms to compute this parameter in trees and series-parallel graphs can also be found in~\cite{Jacobson1989}.

A family of interest for the domination-type parameters are the Cartesian product graphs since the Vizing's conjecture was formulated~\cite{Vizing1968}. This conjecture proposes a general inequality
that relates the domination number of both a Cartesian product graph and its factors. This conjecture is still open and a survey about this subject can be found in~\cite{Bresar2012}, while a recent
new approach is in~\cite{Bresar2021}. Recall that the Cartesian product of two graphs $G\Box H$ is the graph with vertex set $V(G)\times V(H)$ and such that two vertices $(g_1,h_1), (g_2,h_2)$ are adjacent
in $G\Box H$ if either $g_1=g_2$ and $h_1, h_2$ are adjacent in $H$, or $g_1,g_2$ are adjacent in $G$ and $h_1=h_2$. We refer to~\cite{Imrich2000} as a general reference about this topic. It is
well known that domination-type parameters are difficult to handle in Cartesian product graphs and there is no general relationship between the value of such parameters in the product graph and
its factor graphs. Even in the simplest cases of the Cartesian product of two graphs, that is, two paths (grid), a path and a cycle (cylinder) and two cycles (torus) specific procedures are needed
to compute such parameters.

The domination-type parameters in Cartesian product graphs are among the variety of graph parameters that can be computed by using matrix operations. This approach appeared for the first time
in~\cite{Klavzar1996} and has been used in different Cartesian products, such as grids and cylinders, and also in different parameters, such as domination, independent domination and Roman
domination (see for instance~\cite{Crevals2014,Goncalves2011,Guichar2004,Martinez2021,Pavlic2012}). Unlike other parameters, those of domination-type do not use the
usual matrix product but the $(\min,+)$ matrix product, which is also called the tropical product~\cite{Pin1998}. The $(\min, +)$ matrix product is defined over the semi-ring of tropical numbers
$\mathcal{P}=(\mathbb{R}\cup\{\infty\}, \min, +, \infty, 0)$ in the following way: $(A \boxtimes B)_{ij}=\min _k(a_{ik}+b_{kj})$. Moreover, for matrix $A$ and $\alpha\in \mathbb{R}\cup\{\infty\}$,
$(\alpha\boxtimes A)_{ij}=\alpha+a_{ij}$.

Graph algorithms involving tropical algebra operations can be found in literature~\cite{Kepner2011}. The computational side of this approach leads to interesting challenges
bearing in mind the large size of the matrices involved in such algorithms and both, special properties of the matrices and regular structures of the graphs, have been taken into
account in order to reduce the complexity of the matrix computations~\cite{Dobo1990,Felz2011}. Moreover, optimal implementations of the matrix operations in multicore and
GPU platforms have proven to be suitable for these problems~\cite{Buluc2011,Humayun2016,Yang2020}.

A contribution to the problem of the computation of the $2$-domination number in cylinders can be found in~\cite{Garzon2022}, where this parameter was obtained in cylinders with a
small cycle and any path, by using algorithms involving the $(\min,+)$ matrix-vector product. We now focus the complementary problem of computing this parameter in cylinders with a
small path and any cycle, which is unknown. The technique we use here requires performing the $(\min,+)$ matrix-matrix product, which has higher computational requirements.

The goal of this work is twofold. From the computational point of view, efficient routines to compute $(\min,+)$ matrix products on multicore CPUs and GPUs are developed. Moreover, the matrices involved in the analysis of domination-type parameters in graphs are used to evaluate such implementations on modern HPC platforms. It is relevant to underline that, beyond this particular graph analysis, these efficient implementations are useful to accelerate the wide range of applications which are expressed in terms of $(\min,+)$ matrix products. To allow the scientific community to access to these efficient implementations of $(\min,+)$ matrix products, they are available at \href{https://github.com/hpcjmart/2domination}{https://github.com/hpcjmart/2domination}.

From the perspective of the graph analysis, our objective is to conjecture a formula for the $2$-domination number in cylinders with path and cycle of unbounded order. Obtaining the value of the $2$-domination number in cylinders with one small factor, either the path or the cycle, is the first step to addressing the general case. The reason is the regular behavior that is expected, except for the smallest cases. Making such regularity apparent provides the key information to look for the general formula.

In Section~\ref{section:theory} we present the theoretical results that give support to the algorithms shown in Section~\ref{section:algorithms}
along with their computational analysis. Such algorithms will provide the desired values of the $2$-domination number in cylinders with small path and any cycle, which we present in
Section~\ref{section:results}, as well as our conclusions from the computational point of view.

\section{The 2-domination number in cylindrical graphs with small paths}\label{section:theory}

In this section we describe our approach to compute the $2$-domination number of cylinders $P_m\Box C_n$ with small paths. Such approach, involving the $(\min,+)$ matrix-matrix product has also been used to obtain similar results for the Roman domination number~\cite{Martinez2021}. We first describe the general ideas involved in this method and then, we particularize the case of $\gamma_2$.

\subsection{General construction}

We focus on the following result from~\cite{Carre1979}, that we quote from~\cite{Klavzar1996} in the version related to the $(\min,+)$ matrix product.

Let $\mathcal{D}$ be a digraph with vertex set $V(\mathcal{D})= \{v_1, v_2, \dots, v_s\}$ together with a labeling function $\ell$ which assigns an element of the semi-ring $\mathcal{P}=(\mathbb{R}\cup\{\infty\}, \min, +, \infty, 0)$ to every arc of the digraph $\mathcal{D}$.
A \emph{path of length $n$} in $\mathcal{D}$ is a sequence of $n$ consecutive arcs $Q=(v_{i_0}v_{i_1})(v_{i_1}v_{i_2})\dots (v_{i_{k-1}}v_{i_n})$ and $Q$ is \emph{a closed path} if $v_{i_0}=v_{i_{n}}$.
The labeling $\ell$ can be easily extended to paths: $\ell(Q) = \ell(v_{i_0}v_{i_1})+\ell(v_{i_1}v_{i_2})+\dots +\ell(v_{i_{k-1}}v_{i_{n}}).$
\begin{theorem}[\cite{Carre1979}]\label{thm:carre}
Let $S_{ij}^n$ be the set of all paths of length $n$ from $v_i$ to $v_j$ in $\mathcal{D}$ and let $A(\mathcal{D})$ be the
matrix defined by
$$A(\mathcal{D})_{ij} =
\left\{
\begin{array}{ll}
\ell(v_i,v_j) & \text{if }(v_i,v_j) \text{ is an arc of } G,\\
\infty & \text{otherwise.}
\end{array}
\right.
$$
If $A(\mathcal{D})^n$ is the $n$-th $(\min, +)$ power of $A(\mathcal{D})$, then $(A(\mathcal{D})^n)_{ij}=\min \{\ell(Q)\colon Q\in S_{ij}^n\}.$
\end{theorem}
The application of these results to the computation of domination-type parameters in Cartesian product graphs follows a common approach which uses the fact that these kind of parameters are defined as the minimum cardinal of a set having a certain property. We now describe this general procedure.

Let $G$ be a graph and let $a(G)$ be a parameter defined as the minimum cardinal of a vertex subset of $G$ having a certain property $A$. First of all, we have to define a direct graph $\mathcal{D}$
such that there exists a bijective correspondence between the vertex subsets $U\subseteq V(G)$ having the property $A$ and the closed paths $Q$ of $\mathcal{D}$ with fixed length $n$, that we denote by
$U\leftrightarrow Q$. As a second step, we have to define a labeling $\ell$ of the arcs of $\mathcal{D}$ such that if $U \leftrightarrow Q$ then, $\vert U\vert =\ell (Q)$. With such digraph and its associated labeling we can now use Theorem~\ref{thm:carre} to obtain $(A(\mathcal{D})^n)_{ii}  = \min \{\ell(Q)\colon Q\in S_{ii}^n\}=\min\{\vert U \vert \colon U\subseteq V(G) \text{ has property } A, U \leftrightarrow Q, Q\in S_{ii}^n \}.$ That is, the $i-th$ entry $(A(\mathcal{D})^n)_{ii}$ of the main diagonal of the matrix $A(\mathcal{D})^n$ provides the minimum cardinal among all vertex subsets of $G$ having property
$A$ and being identified with closed paths of $\mathcal{D}$ starting and ending in $v_i$. Finally, the minimum entry of the main diagonal of $A(\mathcal{D})^n$ gives the desired value of parameter $a(G)$:
\begin{equation*}
\begin{split}
\min_i (A(\mathcal{D})^n)_{ii}&
=\min_i(\min \{\vert U\vert \colon \!\! U\subseteq V(G) \text{ has property } A, U \leftrightarrow Q, Q\in S_{ii}^n \})\\
& =\min\{\vert U\vert \colon U\subseteq V(G) \text{ has property } A \}= a(G)\\
\end{split}
\end{equation*}

A restriction that occurs when using this approach to compute a parameter $a(G)$ is that graph $G$ needs some structure that allows us to identify the vertex subsets $U\subseteq V(G)$ having
the property $A$ and the closed paths $Q$ of $\mathcal{D}$ with fixed length $n$. The Cartesian products of paths and cycles have such structure, as we now briefly sketch. The cylinder $P_m\Box C_n$ has vertex set
$V(P_m\Box C_n)=\{ u_{ij}\colon 0\leq i\leq m-1, 0\leq j\leq n-1\}$. The $j-th$ column is the subgraph generated by $\{u_{ij}\colon 0\leq i\leq m-1\}$, which is isomorphic to $P_m$.

Let $U\subseteq V(P_m\Box C_n)$ be a vertex subset having the property $A$ and let us consider $U_j$ the $j-th$ column of $P_m\Box C_n$, taking into account whether or not its vertices belong to $U$ (by using a labeling of the vertices).
The vertices of the digraph $\mathcal{D}$ are all possible $U_j$ obtained in such way, for every vertex subset having property $A$. Moreover, there is an arc from $U_r$ to $U_{r+1}$, that is, there is an arc from a vertex of
$\mathcal{D}$ to another one if they are consecutive columns in $P_m\Box C_n$ for the same vertex subset $U$ having property $A$. Then, $U$ can be identified with the closed path $Q=(U_1,U_2), (U_2,U_3)\dots (U_n, U_1)$ that has fixed length $n$.

The key point of the construction above is the column structure of the cylinder $P_m\Box C_n$ and additional requirements are needed in such construction depending on the studied parameter $a(G)$.
In this paper we focus on $2$-domination number $\gamma_2$ of the cylinder $P_m\Box C_n$ and a suitable digraph $\mathcal{D}$ will be defined. The $(\min,+)$ powers of the matrix $A(\mathcal{D})$
have to be computed and this matrix is expected to be quite large, to such an extent as digraph $\mathcal{D}$ is much larger than the cylinder $P_m\Box C_n$. Indeed, the matrix size exponentially grows
with the order of the cylinder and for this reason, this approach is useful just in cylinders $P_m\Box C_n$ with small enough values of both $m$ and $n$. An additional procedure involving well-known
properties of the $(\min, +)$ matrix product allows the removal of one of such size restrictions.

\subsection{Specific construction for the 2-domination number}

Let $P_m\Box C_n$ be a cylinder and let $S\subseteq V(P_m\Box C_n)$ a $2$-dominating set. We label the vertices in the cylinder according to the following rules:
\par\bigskip

\begin{itemize}
\item $v=0$ if $v\in S$,
\item $v=1$ if $v\notin S$ and $v$ has at least $2$ neighbors in $S$ in its column or the previous one,
\item $v=2$ if $v\notin S$ and $v$ has just $1$ neighbor in $S$ in its column or the previous one.
\end{itemize}

We now identify each column with a word $p=(p^1, p^2, \dots, p^m)$ with length $m$ in the alphabet $\{0,1,2\}$ and containing neither the sequences $020, 111, 211, 112, 212$ in any position,
nor the sequences $11, 12$ at the beginning (that is, for the letters $p^1 p^2$) nor the sequences $11,21$ at the end (that is, for the letters $p^{m-1} p^m$). These restrictions come from
the fact that $S$ is a $2$-dominating set and from the definition of the labeling. We call \emph{correct m-words} to words of length $m$ in the alphabet $\{0,1,2\}$ fulfilling all the conditions above.
We define the vertex set of the digraph $\mathcal{D}_m$ as the set of all correct $m$-words.

We now focus on the definition of the arcs in the digraph $\mathcal{D}_m$. Given two correct m-words $p=(p^1, p^2, \dots, p^m)$ and $q=(q^1, q^2, \dots, q^m)$, we say that $p$ can follow a $q$ if
they can be consecutive columns (in the order $qp$) in some $2$-dominating set, that is, they follow the rules of the labeling:
\begin{itemize}
\item if $q_i=2$ then $p_i=0$,
\item if $p_i=2$ then exactly one among $p_{i-1}, p_{i+1}, q_i$ is equal to $0$ (if $i=1$ then exactly one among $p_{i+1}, q_i$ is equal to $0$ and if $i=m$ then exactly one among $p_{i-1}, q_i$ is equal to $0$),
\item if $p_i=1$ then at least two among $p_{i-1}, p_{i+1}, q_i$ is equal to $0$ (the same comment as above for cases $i=1$ and $i=m$).
\end{itemize}

Finally, there is an arc from a word $q$ to a word $p$ if and only if $p$ can follow $q$. This concludes the construction of the digraph $\mathcal{D}_m$, and it is clear that every $2$-dominating
set $S$ of $P_m\Box C_n$ is univocally identified with a closed path $Q$ of length $n$, that is, $S\leftrightarrow Q$.

\begin{figure}[ht]
     \centering
     \begin{subfigure}{.21\textwidth}
     \centering
        \subcaptionbox{{\color{black}The black vertices $2$-dominate $P_4\Box C_5$}\label{fig:example_a}}[3.3cm] {\includegraphics[width=\textwidth]{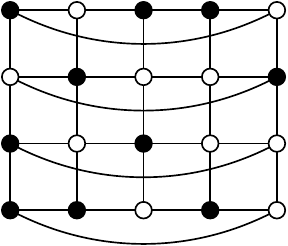}}
     \end{subfigure}
     \hspace{2cm}
     \begin{subfigure}{.2\textwidth}
     \centering
         \subcaptionbox{{\color{black}The vertex labeling provides a word list} \label{fig:example_b}}[3.3cm]
         {\includegraphics[width=\textwidth]{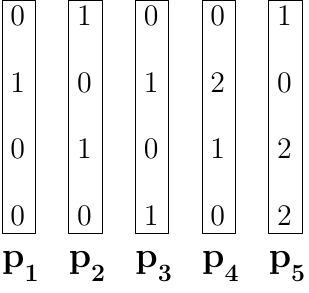}}

     \end{subfigure}
        \caption{{\color{black}A $2$-dominating set of $P_4\Box C_5$ and its associated word list.}}
        \label{fig:example}
\end{figure}

We now need to define a labeling of the arcs of $\mathcal{D}_m$ fulfilling that if $S\leftrightarrow Q$ then, $\vert S \vert =\ell(Q)$. To this end, for an arc $(q, p)$ we define its label as $\ell(q,p)=$number
of zeros of $p$, which obviously gives the desired property. {\color{black} We illustrate the definitions above with an example.

\begin{example}\label{ex:d4}
In Figure~\ref{fig:example_a} a $2$-dominating set of $P_4\Box C_5$ is shown (black vertices). Moreover, the list of correct words representing the columns of such $2$-dominating sets are in Figure~\ref{fig:example_b}.

Clearly $p_{i+1}$ can follow $p_i$ for $i\in \{1,2,3,4\}$ and $p_1$ can follow $p_5$ so $Q=(p_1,p_2), (p_2,p_3), (p_3,p_4), (p_4,p_5),(p_5,p_1)$ is a closed path in the digraph $\mathcal{D}_4$. The label of each arc of $Q$ is the number of zeros in the second word, that is, $\ell(p_1,p_2)=2, \ell(p_2,p_3)=2, \ell(p_3,p_4)=2, \ell(p_4,p_5)=1,\ell(p_5,p_1)=3$. Hence $\ell(Q)=2+2+2+1+3=10$, that reflects that the $2$-dominating set has $10$ vertices.
\end{example}}

\begin{theorem}\label{thm:computation}
Let $P_m\Box C_n$ be a cylinder and let $\mathcal{D}_m$ be the digraph constructed above, with the arc labeling $\ell$. Let $S_{qp}^n$ be the set of all paths of length $n$ from $q$ to $p$
in $\mathcal{D}_m$ and let $A(\mathcal{D}_m)$ be the matrix defined by
$$A(\mathcal{D}_m)_{qp} =
\left\{
\begin{array}{ll}
\ell(q,p) & \text{if }(q,p) \text{ is an arc of } G,\\
\infty & \text{otherwise.}
\end{array}
\right.
$$
If $A(\mathcal{D}_m)^n$ is the $(\min, +)$ power of $A(\mathcal{D}_m)$ then, $\min_i (A(\mathcal{D}_m)^n)_{ii}= \gamma_2(P_m\Box C_n).$
\end{theorem}

\proof
The proof comes from Theorem~\ref{thm:carre} and the specific constructions of the digraph $\mathcal{D}_m$ and the labeling $\ell$. \qed
\par\medskip

{\color{black}Roughly speaking, Theorem~\ref{thm:carre} says that the entry $(i,j)$ of the matrix $A(\mathcal{D}_m)^n$ gives the minimum label among all paths in $\mathcal{D}_m$ with length $n$, beginning in $p_i$ and ending in $p_j$. Therefore, the entry $(i,i)$ on the main diagonal shows the minimum label among all closed $n$-paths that begin and end in $p_i$. Each closed path represents a $2$-dominating set of $P_m\Box C_n$ and its label is the cardinal of such set (see Figure~\ref{fig:example}). Hence, Theorem~\ref{thm:computation} says that the minimum entry of the main diagonal gives the minimum cardinal among all $2$-dominating sets, that is, the $2$-dominating number.}

Using Theorem~\ref{thm:computation} to compute the $2$-domination number of $P_m\Box C_n$ is subject to certain restrictions for both $m$ and $n$. On the one hand, the path order $m$ determines the number of correct m-words and therefore, the size of the matrix $A(\mathcal{D}_m)$ that is expected to be of the order of $3^m$. On the other hand, the cycle order $n$ is the number of $(\min,+)$ matrix powers that have to be computed to obtain the value of the $2$-domination number. The first limitation is intrinsic to this approach. However, there are some properties of the $(\min,+)$ matrix product that can avoid the second one.

\begin{lemma}\label{lem:recurrence}
Let $M$ be a square matrix. Suppose that there exist natural numbers $n_0,a,b$ such that $M^{n_0+a}=b\boxtimes M^{n_0}$. Then, $M^{n+a}=b\boxtimes M^{n}$, for every $n\geq n_0$.
\end{lemma}

\proof
By hypothesis, $M^{n_0+a}=b\boxtimes M^{n_0}$. Let $n\geq n_0$ be such that $M^{n+a}=b\boxtimes M^{n}$ then,
$M^{(n+1)+a}=M\boxtimes M^{n+a}=M\boxtimes(b\boxtimes M^{n})=b\boxtimes(M\boxtimes M^{n})=b\boxtimes M^{n+1}$. \qed

\begin{theorem}\label{thm:complete}
Let $m\geq 2$ be an integer and suppose that there exist natural numbers $n_0,a,b$ such that $A(\mathcal{D}_m)^{n_0+a}=b\boxtimes A(\mathcal{D}_m)^{n_0}$.
Then, the $2$-domination number satisfies the finite difference equation $\gamma_2(P_m\Box C_{n+a})-\gamma_2(P_m\Box C_{n})=b, n\geq n_0$.
\end{theorem}

\proof
By Lemma~\ref{lem:recurrence}, we know that $A(\mathcal{D}_m)^{n+a}=b\boxtimes A(\mathcal{D}_m)^{n}$ for every $n\geq n_0$. Now, by Theorem~\ref{thm:computation} we obtain
$\gamma_2(P_m\Box C_{n+a})  = \min_i (A(\mathcal{D}_m)^{n+a})_{ii} = \min_i (b\boxtimes A(\mathcal{D}_m)^{n})_{ii} =
     b+\min_i (A(\mathcal{D}_m)^n)_{ii}=b+\gamma_2(P_m\Box C_{n})$, for $n\geq n_0$. \qed
\par\medskip

Assuming that $m$ is small enough to apply Theorem~\ref{thm:computation} and that $n_0, a,b$ have been obtained for $m$ then, the boundary values of the finite difference equation above,
that is, $\gamma_2(P_m\Box C_{n})$ for $n_0\leq n \leq n_0+a-1$ can be computed by using Theorem~\ref{thm:computation} and the finite difference equation can be easily solved to obtain the formula
for the $2$-domination number $\gamma_2(P_m\Box C_n)$, for $n\geq n_0$. Moreover, the remaining values $\gamma_2(P_m\Box C_{n})$ for $n<n_0$, if any, can also be computed by Theorem~\ref{thm:computation}.
Thus, if $m$ is small enough to apply Theorem~\ref{thm:computation} and the conditions of Theorem~\ref{thm:complete} hold, then $\gamma_2(P_m\Box C_n)$ can be obtained for any $n\geq 3$.

\section{Algorithms and computational analysis}\label{section:algorithms}
In this section we present the algorithms we have used to compute the $2$-domination number of $P_m\Box C_n$, with $2\leq m\leq 12$ and $n\geq 3$. We also study the performance of such algorithms
in sequential and parallel implementations on a CPU AMD EPYC Rome 7642 with 48 cores and, in addition, on a GPU NVIDIA Tesla V100-PCIE with 32 GB of memory, 80 multiprocessors with 128 cores in each multiprocessor (10240 cores CUDA).

Algorithms from~\ref{code:matrix} to~\ref{code:min_recurrence} come from Theorem~\ref{thm:complete} and they allow us to pose the finite difference equation involving the $2$-domination number of $P_m\Box C_n$, with $m$ small enough. Moreover, Theorem~\ref{thm:computation} provides Algorithm~\ref{code:small} to compute the boundary values of the finite difference equations. Our first target is to obtain the suitable values $a_m,b_m,n^m_0$ to pose such equation for each $m\in \{2,\dots ,12\}$ and first of all, we compute the matrix $A(\mathcal{D}_m)$ in Algorithm~\ref{code:matrix}. In order to obtain the set $\mathcal{C}_m$ of all correct $m$-words, we first obtain all the $m$-element variations of $3$-elements $0,1,2$, with repetition allowed. Then, we select those of them not containing the forbidden sequences of the correct $m$-words.

Algorithm~\ref{code:matrix} is only useful for small values of $m$. As we said before, the size of the matrix $A(\mathcal{D}_m)$ is expected to exponentially grow with $m$, as do the necessary computational resources to get and manage such matrix.
\begin{algorithm}[H]
\setstretch{1}
  \caption{Computation of matrix $A(\mathcal{D}_m)$}
  \begin{algorithmic}[1]
    \Require{$m\geq 2$, $\mathcal{C}_m$}
    \Ensure{matrix $A(\mathcal{D}_m)$}
    \For{each $q_i\in \mathcal{C}_m$} \Comment{$i=1;i\leq \vert \mathcal{C}_m\vert ;i$++}
        \For{each $p_j\in \mathcal{C}_m$} \Comment{$j=1;j\leq \vert \mathcal{C}_m\vert ;j$++}
            \State condition=check $p_j$ can follow $q_i$;
            \If{condition==True}
                \State $A(\mathcal{D}_m)_{(i,j)}=$number of zeros of $p_j$;
            \Else
                \State $A(\mathcal{D}_m)_{(i,j)} = \infty$;
            \EndIf
        \EndFor;
    \EndFor;
  \end{algorithmic}
\label{code:matrix}
\end{algorithm}
In Table~\ref{table:sizes} we show the matrix sizes and the memory requirements in cases $2\leq m\leq 13$, by using $16$ bits arithmetic types of integers. The memory size of the matrix in the case $m=13$ makes it unfeasible to allocate it into the GPU memory, which is the processor we have used to accelerate our algorithms. This is the reason we have analyzed, in this paper, the cases $2\leq m\leq 12$. We have run Algorithm~\ref{code:matrix} in the CPU and it takes $2$ minutes in the larger case $m=12$. This running time is small compared with the following algorithms and moreover, the algorithm does not use any matrix operations whose analysis is our objective. Therefore, we have not parallelized this process and the matrix $A(\mathcal{D}_m)$ is an input data for the remaining algorithms.
\begin{center}

\begin{table}[h!]
\centering
\renewcommand{\arraystretch}{1}
\caption{Size of the matrix $A(\mathcal{D}_m)$ in Algorithm~\ref{code:matrix}}%
\setlength{\tabcolsep}{0pt}
\begin{tabular}
{>{\raggedleft\arraybackslash}p{10pt} >{\raggedleft\arraybackslash}p{60pt} >{\raggedleft\arraybackslash}p{75pt} >{\raggedleft\arraybackslash}p{20pt}|| >{\raggedleft\arraybackslash}p{20pt} >{\raggedleft\arraybackslash}p{60pt} >{\raggedleft\arraybackslash}p{75pt}>{\raggedleft\arraybackslash}p{10pt}}
m & Rows & Memory size & &
m & Rows  & Memory size  &\\
\hline
2& 6& $0.07 KB$ & &8&  1386& $3.67 MB$ &
\\
\hline
3&  15& $0.45 KB$ & &  9 &  3447& $22.67 MB$ &
  \\
\hline
4&  36& $2.53 KB$  & & 10&  8568& $140.02 MB$ &
 \\
\hline
5&  90& $15.82 KB$ & &11& 21294& $0.85 GB$ &
 \\
\hline
6& 225&  $98.88 KB$ & & 12&  52929& $5.22 GB$ &
 \\
\hline
7&  558& $0.69 MB$  & &13& 131562& $32.24 GB$  &  \\
\hline
\end{tabular}
\label{table:sizes}
\end{table}
\end{center}
We now need enough $(\min,+)$ powers of the matrix $A(\mathcal{D}_m)$ in order to look for the recurrence relationship. We obtain the desired powers with Algorithm~\ref{code:powers}.
\begin{algorithm}[H]
\setstretch{1}
  \caption{Computation of $K$ $(\min,+)$ powers of $A(\mathcal{D}_m)$}
  \begin{algorithmic}[1]
    \Require{$m\geq 2 $ , $A(\mathcal{D}_m)$ and $K\geq2$}
    \Ensure{$K$ $(\min,+)$ powers of $A(\mathcal{D}_m)$}
    \State initialize $A(\mathcal{D}_m)$; \Comment{Read file from disk}
    \For{$i=1$; $i<=K$; $i++$ }
        \State $A(\mathcal{D}_m)^i= A(\mathcal{D}_m)\boxtimes  A(\mathcal{D}_m)^{i-1}$; \Comment{$(\min ,+)$ Matrix Product}
    \EndFor;
  \end{algorithmic}
  \label{code:powers}
\end{algorithm}
There exist sufficient but not necessary conditions ensuring that the hypotheses in Theorem~\ref{thm:complete} are true (see~\cite{Spalding1998}). However, such conditions provide a non-minimum value for $n_0$ in the order of the square of the matrix size that is not practical. We have run Algorithm~\ref{code:powers} with $K=50$, which has proven to be enough in cases $2\leq m\leq 12$.

Due to the high requirements to sequentially compute the powers, we have modified this routine in two ways to accelerate it on modern multicore CPU and GPUs. On the one hand, we have used the directives of OpenMP~\cite{OpenMP} to parallelize the $(\min,+)$ matrix multiplication on multicore CPUs.
{\color{black} Specifically, we use the OpenMP directives to accelerate the computation of each product, so the outer loop that iterates through the rows of the first matrix of the product is parallelized. This technique is straightforward, and it allows to efficiently develop the $(\min, +)$ matrix product to leverage the resources of the CPU multicore processors. Moreover, the performance achieved is enough for the purpose of our work when the dimensions of the matrices are moderated.
}

On the other hand, the powers have also been carried out by a modification of the routine MatrixMul, available in the NVIDIA CUDA TOOLKIT 11~\cite{nvidia2020} and described in the CUDA C Programming Guide (see~\cite{nvidia2021}, Chapter 3), to adapt it to the $(\min,+)$ multiplication.
{\color{black} In this case, we use a different parallelization strategy than the one used in OpenMP. It based on a tiled matrix multiplication to optimize the GPU hierarchy memory management. So, this method takes advantage of the lower latency, the higher bandwidth shared memory within GPU thread blocks and the number of slow accesses to memory device, which are minimized. For details of the memory access pattern of MatrixMul see Chapter 3 of~\cite{nvidia2021}.}

We show in Table~\ref{table:time} the running times of Algorithm~\ref{code:powers} in cases $7\leq m\leq 12$ while in the remaining cases the algorithm needs less than $1$ second, even with the sequential implementation.
\begin{center}
\vspace{-0.5cm}
\begin{table}[ht]
\renewcommand{\arraystretch}{1}
\centering
\caption{Running times of Algorithm~\ref{code:powers} to compute $A(\mathcal{D}_m)^k, k\leq 50$}%
\setlength{\tabcolsep}{0pt}
\begin{tabular}{>{\raggedright\arraybackslash}p{10pt}> {\raggedleft\arraybackslash}p{60pt}>{\raggedleft\arraybackslash}p{65pt}>{\raggedleft\arraybackslash}p{70pt}>
{\raggedleft\arraybackslash}p{60pt}>{\raggedleft\arraybackslash}p{70pt}}
m & \multicolumn{1}{c}{sequential}   & \multicolumn{2}{c}{multicore 48 threads} &  \multicolumn{2}{c}{GPU}   \\
\cmidrule(lr){2-2} \cmidrule(lr){3-4} \cmidrule(lr){5-6}
 &{\scriptsize{time}} & {\scriptsize{time\ \ \ }}  &  {\scriptsize{sequ./multicore speedup}} &  {\scriptsize{time\ \ \ }} & {\scriptsize{multicore/GPU speedup}} \\
\hline
7& $13.4 s$ & $0.4 s$ & $33.5$ & $0.2 s$ &  $2$\\
\hline
8& $3 m$ $18.9 s$ & $5.3 s$ & $37.5$ & $0.5 s$ & $10.6$\\
\hline
9& $56 m$ $42.3 s$ & $1m$ $31.8 s$  & $37.1$ & $2.9 s$ & $31.7$\\
\hline
10& $17 h$ $9 m$ $21.6 s$ & $25 m$ $23.4 s$ & $40.5$ & $30.2 s$ & $50.4$\\
\hline
11&  \rule{1.5cm}{0.5pt} & $6 h$ $29 m$ $28.5 s$ & & $6 m$ $21.6 s$ & $61.2$\\
\hline
12&  \rule{1.5cm}{0.5pt} & \rule{1.5cm}{0.5pt} & & $1 h$ $30 m$ $15.6 s$ & \\
\hline
\end{tabular}
\label{table:time}
\end{table}
\vspace{-0.5cm}
\end{center}
Table~\ref{table:time} shows that the running time of computing $50$ $(\min,+)$ powers of matrix $A(\mathcal{D}_m)$ exponentially grows as the matrix size increases. In order to address large cases in reasonable time we have run an OpenMP parallel implementation with $48$ cores/threads. Such implementation provides small running times in cases $m=8$ and $m=9$ but it grows fast for $m\geq 10$. In order to increase the efficiency of this algorithm, we have run a version of the $(\min,+)$ matrix product in CUDA for NVIDIA GPU and we have obtained a significant improvement in terms of running times compared to the sequential and the parallel OpenMP versions.

The following step to apply Theorem~\ref{thm:complete} is to find the appropriate recurrence relationship between two powers of matrix $A(\mathcal{D}_m)$. Even though such matrix is sparse, we have noted that its powers become dense, that is, with no infinite entries, from the third one. Therefore, the hypothesis in Theorem~\ref{thm:complete}, that is, $A(\mathcal{D}_m)^{n_0+a}=b\boxtimes A(\mathcal{D}_m)^{n_0}$ is equivalent to $A(\mathcal{D}_m)^{n_0+a}-A(\mathcal{D}_m)^{n_0}$ being a constant matrix with entries equal to $b_m$. We use this fact in Algorithm~\ref{code:recurrence}. The results are shown in Table~\ref{table:no_min_values} ($K=50$).
\vspace{-0.2cm}
\begin{algorithm}[H]
\setstretch{0.78}
  \caption{Search for the recurrence relation}
  \begin{algorithmic}[1]
    \Require{$m\geq 2$ and $K$ $(\min,+)$ powers of $A(\mathcal{D}_m)$}
    \Ensure{$a_m, b_m, r_0^m$}
    \For{$i=K$; $i>=1$; $i- -$ }
        \State $A(\mathcal{D}_m)^{i}$;
        \For{$j=i-1$; $j>=1$; $j- -$ }
        \State initialize $A(\mathcal{D}_m)^{j}$;
        \If{$(A(\mathcal{D}_m)^{i}-A(\mathcal{D}_m)^{j})==constant$ }
            \State $a_m=i-j$; \Comment{Steep}
            \State $b_m=constant$ \Comment{Difference value}
            \State $r^m_0=j$;
            \State Exit;
        \EndIf;
        \EndFor;
    \EndFor;
  \end{algorithmic}
  \label{code:recurrence}
\end{algorithm}
\begin{center}
\begin{table}[ht]
\renewcommand{\arraystretch}{0.9}
\centering
\caption{Results obtained by Algorithm~\ref{code:recurrence}}%
\setlength{\tabcolsep}{3pt}
\begin{tabular}{
>{\raggedleft\arraybackslash}p{10pt}>{\raggedleft\arraybackslash}p{25pt}
>{\raggedleft\arraybackslash}p{25pt} >{\raggedleft\arraybackslash}p{25pt}>{\raggedleft\arraybackslash}p{10pt}|| >{\raggedleft\arraybackslash}p{10pt}>{\raggedleft\arraybackslash}p{25pt}
>{\raggedleft\arraybackslash}p{25pt} >{\raggedleft\arraybackslash}p{25pt}
}
m  &$r^m_0$ & $a_m$ & $b_m$ & & m  &$r^m_0$ & $a_m$ & $b_m$\\
\hline
2& 48& 2 &2 && 8& 47 & 3 & 10 \\
\hline
3& 44& 6 &8  && 9& 47 & 3 & 11\\
\hline
4& 42& 8 &14 && 10& 47 & 3 & 12 \\
\hline
5&  43& 7& 15 & &11& 47 & 3 & 13 \\
\hline
6&  39 & 11 & 28 && 12& 47 & 3 & 14 \\
\hline
7&  32 & 18 & 53 && & & & \\
\hline
\end{tabular}
\label{table:no_min_values}
\end{table}
\end{center}
It is expected that the values of $r^m_0$ are not minimum because we have found a recurrence relationship with $r^m_0+a_m=50$, for every $m$. But in any case, we have confirmed that matrix
$A(\mathcal{D}_m)$ meets the hypothesis of Theorem~\ref{thm:complete} and the finite difference equation can be posed for $n\geq r^m_0$.

We now show how to obtain the minimum value $n^m_0$ such that $A(\mathcal{D} _m)^{n+a_m}=b_m\boxtimes A(\mathcal{D}_m)^{n}$ for every $n\geq n^m_0$, in Algorithm~\ref{code:min_recurrence} . Finding this optimal value could be interesting in order to try to reduce the number of $(\min,+)$ powers required to ensure the hypothesis of Theorem~\ref{thm:complete}.
\begin{algorithm}[H]
\setstretch{0.8}
  \caption{Search for the minimum values of the recurrence relationship}
  \begin{algorithmic}[1]
    \Require{$m\geq 2$ and $r_0^m, a_m, b_m$}
    \Ensure{$n_0^m$}
    \For{$i=(r_0^m+a_m)$; $i>=1$; $i- -$}
        \If{$(A(\mathcal{D}_m)^{i}-A(\mathcal{D}_m)^{i-a_m})\neq b_m$ }
            \State Exit;
        \Else
            \State $n_0^m=i-a_m$;
        \EndIf;
    \EndFor;
   \end{algorithmic}
  \label{code:min_recurrence}
\end{algorithm}
We show the values of $n^m_0$ obtained with Algorithm~\ref{code:min_recurrence} in Table~\ref{table:values}, together with the values of $a_m, b_m$ shown before. Such values provide the finite difference equation  $\gamma_2(P_m\Box C_{n+a_m})-\gamma_2(P_m\Box C_{n})=b_m, n\geq n^m_0$ and $m\in\{2,\dots,12\}$.
\begin{center}
\begin{table}[ht]
\renewcommand{\arraystretch}{1}
\centering
\caption{Values to apply Theorem~\ref{thm:complete} obtained with Algorithm~\ref{code:min_recurrence}}%
\setlength{\tabcolsep}{3pt}
\begin{tabular}{
>{\raggedleft\arraybackslash}p{10pt}>{\raggedleft\arraybackslash}p{25pt}
>{\raggedleft\arraybackslash}p{25pt} >{\raggedleft\arraybackslash}p{25pt} >{\raggedleft\arraybackslash}p{10pt} ||
>{\raggedleft\arraybackslash}p{10pt}>{\raggedleft\arraybackslash}p{25pt}
>{\raggedleft\arraybackslash}p{25pt} >{\raggedleft\arraybackslash}p{25pt}
}
m  &$n^m_0$ & $a_m$ & $b_m$ & & m  &$n^m_0$ & $a_m$ & $b_m$  \\
\hline
2& 4& 2 &2 && 8& 25 & 3 & 10\\
\hline
3& 7& 6 &8 && 9& 22 & 3& 11\\
\hline
4& 9& 8 &14 && 10& 21 & 3 & 12 \\
\hline
5&  31& 7& 15&& 11& 24 & 3 & 13 \\
\hline
6&  19 & 11 & 28 && 12& 26 & 3 & 14\\
\hline
7&  23 & 18 & 53 && & & & \\
\hline
\end{tabular}
\label{table:values}
\end{table}
\end{center}
The matrix operation used in Algorithms~\ref{code:recurrence} and~\ref{code:min_recurrence} is the matrix difference, which consumes fewer computational resources than the $(\min,+)$ matrix multiplication.
{\color{black} Indeed, both algorithms are faster with the OpenMP directives than on the GPU due to the cost of communications to allocate the matrices on the GPU memory to perform quite a simple operation. For instance, the running times (in seconds) of Algorithm~\ref{code:recurrence} for largest case we have computed $m=12$ are  $16.8$ on the CPU (sequential), $ 13.6$ on the GPU and $7.2$ with OpenMP (48 cores). For Algorithm~\ref{code:min_recurrence}, they are $149.8$, $170.0$ and $98.5$, respectively.}

Finally, we compute the boundary values needed to solve the finite difference equations and to obtain the formul\ae\ of the $2$-domination number in the studied cases, with Algorithm~\ref{code:small}, by using Theorem~\ref{thm:computation}.
\begin{algorithm}[H]
\setstretch{1}
  \caption{$\gamma _2(P_m\Box C_n)$, for $3\leq n\leq n_0^m+a_m-1$}
  \begin{algorithmic}[1]
    \Require{$m\geq 2$ and $(n_0^m+a_m-1)$ $(\min,+)$ powers of $A(\mathcal{D}_m)$}
    \Ensure{$\gamma_2(P_m\Box C_n), 3\leq n \leq (n_0^m+a_m-1)$}
    \For{$i=3$; $i<=(n_0^m+a_m-1)$; $i++$}
        \State $\min$(main diagonal($A(\mathcal{D}_m)^i$))
    \EndFor
  \end{algorithmic}
  \label{code:small}
\end{algorithm}
Algorithm~\ref{code:small} uses the minimization operation over the main diagonal of the matrix $A(\mathcal{D}_m)^i$, which can be seen as a vector with a length of the number of rows of the matrix. This matrix operation is less computationally demanding given that the number of the operations needed here is on the order of the number of rows of the matrix while in Algorithms~\ref{code:recurrence} and~\ref{code:min_recurrence} the order is the square of that number. Indeed, the CPU needs less than $1$ second if $m\leq 11$ and $11.8$ seconds in the largest case $m=12$. Our program to compute the $2$-domination number of cylindrical graphs with small paths consists of consecutive run Algorithms from~\ref{code:powers} to~\ref{code:small} and we have implemented it in four ways. The first one runs every algorithm on the CPU and we have here completed the computation of cases $m\leq 10$, due to high running times of Algorithm~\ref{code:powers}.

In the second version we have used the OpenMP directives to parallelize the execution of the $(\min,+)$ matrix product routines in Algorithm~\ref{code:powers} and the matrix difference in Algorithm~\ref{code:recurrence} and~\ref{code:min_recurrence} because they are the most computationally demanding matrix operations. We have computed until case $m=11$ with 48 cores and although the speedup for Algorithm~\ref{code:powers} is over 40 in the last case, the running time is still huge. The third program runs Algorithms~\ref{code:powers}, \ref{code:recurrence} and~\ref{code:min_recurrence} on the GPU and cases $m\leq 12$ have been obtained. Algorithm~\ref{code:powers} presents here a very noticeable improvement in terms of running time, but the huge matrix size does not allow us to approach large cases given that from $m=13$ the matrix can not be allocated on the GPU memory.

In order to test the goodness of the implementation of Algorithms~\ref{code:recurrence} and~\ref{code:min_recurrence} on the CPU compared to the GPU, we have done the fourth version that uses the GPU just in Algorithm~\ref{code:powers} and the OpenMP parallelization for Algorithms~\ref{code:recurrence} and~\ref{code:min_recurrence}. This is slightly faster than version 3 because of the communication costs to allocate matrices on the GPU memory to perform matrix operations with little computational cost. The total running times of the four versions are shown in Table~\ref{table:total_time}.
\begin{center}
\begin{table}[ht]
\renewcommand{\arraystretch}{1}
\centering
\caption{Total running times}%
\setlength{\tabcolsep}{3pt}
\begin{tabular}{>{\raggedleft\arraybackslash}p{10pt}>{\raggedleft\arraybackslash}p{50pt}
>{\raggedleft\arraybackslash}p{50pt} >{\raggedleft\arraybackslash}p{50pt}>{\raggedleft\arraybackslash}p{50pt}}
m  & Version 1 & Version 2 & Version 3 &  Version 4 \\
\hline
7&$13.5 s$ &$0.4 s$ & $0.2 s$& $0.3 s$ \\
\hline
8& $3 m$ $19 s$ & $5.4 s$ & $0.6 s$ & $0.6 s$\\
\hline
9&$56 m $ $43 s$ & $1 m $ $32 s$ &$3.9 s$ & $3.5 s$\\
\hline
10& $17 h$ $9 m$  $27 s$ &$25 m$ $26 s$ & $35 s$ & $32.5 s$\\
\hline
11&  & $6 h$ $29 m$ $45 s$ & $6 m $ $44 s$ & $6m $ $32 s$\\
\hline
12& & & $1 h$ $33 m$  $34 s$ & $1 h$ $29 m$ $58 s$\\
\hline
\end{tabular}
\label{table:total_time}
\end{table}
\end{center}

\section{Conclusions}\label{section:results}
According to Theorem~\ref{thm:complete}, values in Table~\ref{table:values}
allow us to pose the finite difference equation $\gamma_2(P_m\Box C_{n+a_m})-\gamma_2(P_m\Box C_n)=b_m, n\geq n^m_0$, for each $2\leq m\leq 12$. The boundary values $\gamma_2(P_m\Box C_{n})$, $n^m_0\leq n\leq n^m_0+a_m-1$, have been obtained with Algorithm~\ref{code:small}. Therefore, the solution is $\gamma_2(P_m\Box C_n)=\left\lceil \frac{b_m \cdot n}{a_n}\right\rceil+\alpha^m_k $, where $n\equiv k\pmod {a_m}$ and $\alpha^m_k$ depends on the boundary values for each $m$. Moreover, the remaining values of $\gamma_2(P_m\Box C_n)$, for $3\leq n<n^m_0$,
have also been computed with Algorithm~\ref{code:small} and most of them follow the general formula.

In the same way as in other domination parameters in grids and cylinders (see~\cite{Crevals2014,Goncalves2011}), these results show a non-regular behavior for the smallest values of $m$, but it becomes regular for $m\geq 8$. Note that if $8\leq m\leq 12$ then, $a_m=3$ and $b_m=m+2$.
In such cases $\gamma_2(P_m\Box C_n)=\lceil \frac{(m+2)n}{3}\rceil+\alpha^m_k$,
where $n\equiv k\pmod 3$ and $\alpha^m_k$ again depends on the boundary values $\gamma_2(P_m\Box C_{n^m_0+k})$. In order to complete the formul\ae, in Table~\ref{table:restos} we show the values of $\alpha_k^m$, for each $m\in \{2, \dots ,12\}$ and $k\in \{0,\dots ,a_m-1\}$.

The only exceptions are $n=5$, for $m\in\{8, 10,12\}$, where  $\alpha_k^m =2$. This value is coherent with the results obtained in~\cite{Garzon2022}: $\gamma_2(C_5\Box P_{m})=2m+2$ if $2<m\equiv 0\pmod 2$ and $\gamma_2(C_5\Box P_{m})=2m+1$ if $m=2$ or $m\equiv 1\pmod 2$.

\begin{center}
\begin{table}[ht]
\renewcommand{\arraystretch}{1}
\centering
\caption{Values of $\alpha_k^m$}%
\setlength{\tabcolsep}{3pt}
\begin{tabular}{>{\raggedright\arraybackslash}p{14pt}>{\raggedleft\arraybackslash}p{20pt}
>{\raggedleft\arraybackslash}p{130pt}>{\raggedleft\arraybackslash}p{60pt} }
m  &$a_m$ & $k$ with $\alpha_k^m =1$& $k$ with $\alpha_k^m =0$ \\
\hline
2& 2 & none & all \\
\hline
3& 6 & none & all\\
\hline
4& 8 & 4,5 & otherwise \\
\hline
5& 7 & none & all \\
\hline
6& 11 & 5,9 & otherwise \\
\hline
7&  18 & $k\in \{1,2,4,5\}$, $19\leq n\equiv k\pmod {18}$ & otherwise\\
\hline
8$^*$& 3 & none & all\\
\hline
9& 3 & none & all  \\
\hline
10$^*$ & 3 & 1,2 & otherwise\\
\hline
11& 3 & 2 & otherwise\\
\hline
12$^*$& 3 &  1,2  & otherwise  \\
\hline
\multicolumn{4}{l}{$^*$ There is one exception}
\end{tabular}
\label{table:restos}
\end{table}
\end{center}

In spite of obtaining that $\alpha_k^m\leq 2$ for $m\leq 12$, we think that such numbers will increase for some values of $n$ as $m$ grows because they would depend on $m$ in some way. Our results cover the cases $2\leq m\leq 12$, $3\leq n\leq 15$ already studied in~\cite{Garzon2022}, and all the results match. In addition, for $8\leq m\leq 12$ and $n\equiv 0\pmod 3$ we have shown that $\gamma_2(P_m\Box C_n)=\frac{(m+2)n}{3}$. The same formula for $n=3,6,9,12,15$ and $m\geq 8$ is obtained in~\cite{Garzon2022} and we have now extended this result to every $n\equiv 0\pmod 3$, for $8\leq m\leq 12$. Also note that our formul\ae \ for $m\leq 7$ and $n\equiv 0\pmod 3$ show that such small cases do not follow the same formula, in general. Our results together with those in~\cite{Garzon2022} give us support to conjecture that $\gamma_2(P_m\Box C_n)=\frac{(m+2)n}{3}$, if $m\geq 8$ and $n\equiv 0\pmod 3$.

Regarding the computational point of view, our main target was to develop efficient routines to compute $(\min,+)$ matrix products on multicore CPUs and GPUs. Such routines have application to the computation of the $2$-domination number of cylindrical graphs with small paths of order $m$. Our approach has as a limitation the size of the involved matrices that exponentially grows as $m$ does. This condition has led us to focus on cases $2\leq m\leq 12$ that meet the requirements of our computational resources on both the CPU and the GPU.

Once we have obtained the matrices for cases $2\leq m\leq 12$, we have divided the routines in Algorithms from~\ref{code:powers} to~\ref{code:small} and three of them, Algorithms~\ref{code:recurrence}, \ref{code:min_recurrence} and~\ref{code:small}, can be run on the CPU in a reasonable time. Moreover, the OpenMP parallelization with 48 cores slightly improves such running times, which are negligible compared to the total ones. However, the CPU has shown to be non sufficient to run Algorithm~\ref{code:powers} in the most interesting cases, which are the largest ones, to find the desired regular behavior of the $2$-domination number. The matrix operation used by this algorithm is the $(\min,+)$ matrix product and we explore two improvement options to reduce its running time: a parallelization of the algorithm with OpenMP with 48 cores and an implementation of this matrix product in CUDA for NVIDIA GPU. The OpenMP parallel version with 48 cores of Algorithm~\ref{code:powers} has shown a speedup over $40$ regarding the sequential version in case $m=10$. However, the running times are so high that the parallelization is not enough for $m\geq 11$, where more than $6$ hours are needed. In contrast, the GPU version computes $50$ powers of the matrix $A(\mathcal{D}_m)$ in considerably less time, with a speedup over $60$ compared to the OpenMP version for $m=12$.

We think it would be possible to improve the efficiency of Algorithm~\ref{code:powers} by reducing the number of computed powers while the finite difference equation can still be solved. In addition, some parallelization of the $(\min,+)$ product allowing to distribute the product of two matrices in small sets of rows and columns would give the opportunity of computing some cases larger than $m=12$. Such improvements  would perhaps allow us to conjecture a general formula of the $2$-domination number of the cylinder $P_m\Box C_n$ with $n\equiv 1, 2 \pmod 3$.

To sum up, we have solved the graph problem of computing the $2$-domination number of some cylinders with a small path in a reasonable time by exploiting the benefits of the GPU's to run algorithms involving the $(\min,+)$ matrix product while the rest of matrix operations involved, such as the matrix difference or the minimization of the main diagonal of a matrix, demand fewer computational resources and they can be addressed on the multicore CPU in a short time. Finally, we have conjectured that $\gamma_2(P_m\Box C_n)$ if $n\equiv 0\pmod 3$.

\backmatter
\bmhead{Acknowledgments}
These results are part of the projects RTI2018-095993-B-I00 and PID2019-104129GB-I00 both funded by MCIN/AEI/10.13039/501100011033/ FEDER ``A way to make Europe.''

\bibliography{bibfile}


\begin{thebibliography}{33}
\ifx \bisbn   \undefined \def \bisbn  #1{ISBN #1}\fi
\ifx \binits  \undefined \def \binits#1{#1}\fi
\ifx \bauthor  \undefined \def \bauthor#1{#1}\fi
\ifx \batitle  \undefined \def \batitle#1{#1}\fi
\ifx \bjtitle  \undefined \def \bjtitle#1{#1}\fi
\ifx \bvolume  \undefined \def \bvolume#1{\textbf{#1}}\fi
\ifx \byear  \undefined \def \byear#1{#1}\fi
\ifx \bissue  \undefined \def \bissue#1{#1}\fi
\ifx \bfpage  \undefined \def \bfpage#1{#1}\fi
\ifx \blpage  \undefined \def \blpage #1{#1}\fi
\ifx \burl  \undefined \def \burl#1{\textsf{#1}}\fi
\ifx \doiurl  \undefined \def \doiurl#1{\url{https://doi.org/#1}}\fi
\ifx \betal  \undefined \def \betal{\textit{et al.}}\fi
\ifx \binstitute  \undefined \def \binstitute#1{#1}\fi
\ifx \binstitutionaled  \undefined \def \binstitutionaled#1{#1}\fi
\ifx \bctitle  \undefined \def \bctitle#1{#1}\fi
\ifx \beditor  \undefined \def \beditor#1{#1}\fi
\ifx \bpublisher  \undefined \def \bpublisher#1{#1}\fi
\ifx \bbtitle  \undefined \def \bbtitle#1{#1}\fi
\ifx \bedition  \undefined \def \bedition#1{#1}\fi
\ifx \bseriesno  \undefined \def \bseriesno#1{#1}\fi
\ifx \blocation  \undefined \def \blocation#1{#1}\fi
\ifx \bsertitle  \undefined \def \bsertitle#1{#1}\fi
\ifx \bsnm \undefined \def \bsnm#1{#1}\fi
\ifx \bsuffix \undefined \def \bsuffix#1{#1}\fi
\ifx \bparticle \undefined \def \bparticle#1{#1}\fi
\ifx \barticle \undefined \def \barticle#1{#1}\fi
\bibcommenthead
\ifx \bconfdate \undefined \def \bconfdate #1{#1}\fi
\ifx \botherref \undefined \def \botherref #1{#1}\fi
\ifx \url \undefined \def \url#1{\textsf{#1}}\fi
\ifx \bchapter \undefined \def \bchapter#1{#1}\fi
\ifx \bbook \undefined \def \bbook#1{#1}\fi
\ifx \bcomment \undefined \def \bcomment#1{#1}\fi
\ifx \oauthor \undefined \def \oauthor#1{#1}\fi
\ifx \citeauthoryear \undefined \def \citeauthoryear#1{#1}\fi
\ifx \endbibitem  \undefined \def \endbibitem {}\fi
\ifx \bconflocation  \undefined \def \bconflocation#1{#1}\fi
\ifx \arxivurl  \undefined \def \arxivurl#1{\textsf{#1}}\fi
\csname PreBibitemsHook\endcsname

\bibitem{Pin1998}
\begin{bchapter}
\bauthor{\bsnm{Pin}, \binits{J.-E.}}:
\bctitle{2}.
\bbtitle{Tropical semirings,\,}.
\bsertitle{Idempotency (J. Gunawardena, Ed.) Publications of the Newton Institute},
pp. \bfpage{50}--\blpage{69}.
\bpublisher{Cambridge University Press},
\blocation{Cambridge, UK}
(\byear{1998}).
\doiurl{10.1017/CBO9780511662508.004}
\end{bchapter}
\endbibitem

\bibitem{Omanovic21}
\begin{barticle}
\bauthor{\bsnm{Omanovic}, \binits{A.}},
\bauthor{\bsnm{Kazan}, \binits{H.}},
\bauthor{\bsnm{Oblak}, \binits{P.}},
\bauthor{\bsnm{Curk}, \binits{T.}}:
\batitle{Sparse data embedding and prediction by tropical matrix factorization}.
\bjtitle{{BMC} Bioinform.}
\bvolume{22}(\bissue{1}),
\bfpage{89}
(\byear{2021})
\end{barticle}
\endbibitem

\bibitem{Speyer2009}
\begin{barticle}
\bauthor{\bsnm{Speyer}, \binits{D.}},
\bauthor{\bsnm{Sturmfels}, \binits{B.}}:
\batitle{Tropical mathematics}.
\bjtitle{Math. Mag.}
\bvolume{82}(\bissue{3}),
\bfpage{163}--\blpage{173}
(\byear{2009}).
\doiurl{10.1080/0025570X.2009.11953615}
\end{barticle}
\endbibitem

\bibitem{Klavzar1996}
\begin{barticle}
\bauthor{\bsnm{Klav\v{z}ar}, \binits{S.}},
\bauthor{\bsnm{\v{Z}erovnik}, \binits{J.}}:
\batitle{Algebraic approach to fasciagraphs and rotagraphs}.
\bjtitle{Discret. Appl. Math.}
\bvolume{68}(\bissue{1}),
\bfpage{93}--\blpage{100}
(\byear{1996}).
\doiurl{10.1016/0166-218X(95)00058-Y}
\end{barticle}
\endbibitem

\bibitem{Butkovic2019}
\begin{barticle}
\bauthor{\bsnm{Butkovi\v{c}}, \binits{P.}}:
\batitle{A note on tropical linear and integer programs}.
\bjtitle{J. Optim. Theory Appl.}
\bvolume{180}(\bissue{3}),
\bfpage{1011}--\blpage{1026}
(\byear{2019}).
\doiurl{10.1007/s10957-018-1429-8}
\end{barticle}
\endbibitem

\bibitem{Krivulin2015}
\begin{barticle}
\bauthor{\bsnm{Krivulin}, \binits{N.}}:
\batitle{Algebraic solutions of tropical optimization problems}.
\bjtitle{Lobachevskii J. Math.}
\bvolume{36}(\bissue{4}),
\bfpage{363}--\blpage{374}
(\byear{2015}).
\doiurl{10.1134/S199508021504006X}
\end{barticle}
\endbibitem

\bibitem{Haynes1998}
\begin{bbook}
\bauthor{\bsnm{Haynes}, \binits{T.W.}},
\bauthor{\bsnm{Hedetniemi}, \binits{S.T.}},
\bauthor{\bsnm{Slater}, \binits{P.J.}}:
\bbtitle{Fundamentals of Domination in Graphs}.
\bsertitle{Chapman and Hall CRC Pure and Applied Mathematics Series}.
\bpublisher{Marcel Dekker, Inc.},
\blocation{New York, USA}
(\byear{1998})
\end{bbook}
\endbibitem

\bibitem{Fink1985}
\begin{bbook}
\bauthor{\bsnm{Fink}, \binits{J.F.}},
\bauthor{\bsnm{Jacobson}, \binits{M.S.}}:
\bbtitle{N-Domination in Graphs}.
\bsertitle{Graph Theory with Applications to Algorithms and Computer Science},
pp. \bfpage{283}--\blpage{300}.
\bpublisher{John Wiley \& Sons, Inc.},
\blocation{USA}
(\byear{1985})
\end{bbook}
\endbibitem

\bibitem{Butjas2018}
\begin{barticle}
\bauthor{\bsnm{Bujtás}, \binits{C.}},
\bauthor{\bsnm{Jaskó}, \binits{S.}}:
\batitle{Bounds on the 2-domination number}.
\bjtitle{Discrete Appl. Math.}
\bvolume{242},
\bfpage{4}--\blpage{15}
(\byear{2018}).
\doiurl{10.1016/j.dam.2017.05.014}
\end{barticle}
\endbibitem

\bibitem{Garey1979}
\begin{bbook}
\bauthor{\bsnm{Garey}, \binits{M.R.}},
\bauthor{\bsnm{Johnson}, \binits{D.S.}}:
\bbtitle{Computers and Intractability: {A} Guide to the Theory of NP-Completeness}.
\bpublisher{W. H. Freeman},
\blocation{New York, USA}
(\byear{1979})
\end{bbook}
\endbibitem

\bibitem{Jacobson1989}
\begin{barticle}
\bauthor{\bsnm{Jacobson}, \binits{M.S.}},
\bauthor{\bsnm{Peters}, \binits{K.}}:
\batitle{Complexity questions for n-domination and related parameters}.
\bjtitle{Congr. Numer.}
\bvolume{68},
\bfpage{7}--\blpage{22}
(\byear{1989})
\end{barticle}
\endbibitem

\bibitem{Bean1994}
\begin{barticle}
\bauthor{\bsnm{Bean}, \binits{T.J.}},
\bauthor{\bsnm{Henning}, \binits{M.}},
\bauthor{\bsnm{Swart}, \binits{H.C.}}:
\batitle{On the integrity of distance domination in graphs}.
\bjtitle{Australas. J Comb.}
\bvolume{10},
\bfpage{29}--\blpage{44}
(\byear{1994})
\end{barticle}
\endbibitem

\bibitem{Vizing1968}
\begin{barticle}
\bauthor{\bsnm{Vizing}, \binits{V.G.}}:
\batitle{Some unsolved problems in graph theory}.
\bjtitle{Uspekhi Mat. Nauk}
\bvolume{23}(\bissue{6}),
\bfpage{117}--\blpage{134}
(\byear{1968})
\end{barticle}
\endbibitem

\bibitem{Bresar2012}
\begin{barticle}
\bauthor{\bsnm{Bre\v{s}ar}, \binits{B.}},
\bauthor{\bsnm{Dorbec}, \binits{P.}},
\bauthor{\bsnm{Goddard}, \binits{W.}},
\bauthor{\bsnm{Hartnell}, \binits{B.L.}},
\bauthor{\bsnm{Henning}, \binits{M.A.}},
\bauthor{\bsnm{Klav\v{z}ar}, \binits{S.}},
\bauthor{\bsnm{Rall}, \binits{D.F.}}:
\batitle{Vizing's conjecture: a survey and recent results}.
\bjtitle{Journal of Graph Theory}
\bvolume{69}(\bissue{1}),
\bfpage{46}--\blpage{76}
(\byear{2012}).
\doiurl{10.1002/jgt.20565}
\end{barticle}
\endbibitem

\bibitem{Bresar2021}
\begin{barticle}
\bauthor{\bsnm{Bre\v{s}ar}, \binits{B.}},
\bauthor{\bsnm{Hartnell}, \binits{B.L.}},
\bauthor{\bsnm{Henning}, \binits{M.A.}},
\bauthor{\bsnm{Kuenzel}, \binits{K.}},
\bauthor{\bsnm{Rall}, \binits{D.F.}}:
\batitle{A new framework to approach \uppercase{V}izing’s conjecture}.
\bjtitle{Discuss. Math. Graph Theory}
\bvolume{41}(\bissue{3}),
\bfpage{749}--\blpage{762}
(\byear{2021}).
\doiurl{10.7151/dmgt.2293}
\end{barticle}
\endbibitem

\bibitem{Imrich2000}
\begin{bbook}
\bauthor{\bsnm{Imrich}, \binits{W.}},
\bauthor{\bsnm{Klav\v{z}ar}, \binits{S.}}:
\bbtitle{Product Graphs, Structure and Recognition}.
\bsertitle{Wiley-Interscience series in discrete mathematics and optimization.\,},
p. \bfpage{358}.
\bpublisher{Wiley},
\blocation{New York, USA}
(\byear{2000})
\end{bbook}
\endbibitem

\bibitem{Crevals2014}
\begin{barticle}
\bauthor{\bsnm{Crevals}, \binits{S.}}:
\batitle{Domination of cylinder graphs}.
\bjtitle{Congr. Numer.}
\bvolume{219},
\bfpage{53}--\blpage{63}
(\byear{2014})
\end{barticle}
\endbibitem

\bibitem{Goncalves2011}
\begin{barticle}
\bauthor{\bsnm{Gon{\c{c}}alves}, \binits{D.}},
\bauthor{\bsnm{Pinlou}, \binits{A.}},
\bauthor{\bsnm{Rao}, \binits{M.}},
\bauthor{\bsnm{Thomass{\'{e}}}, \binits{S.}}:
\batitle{The domination number of grids}.
\bjtitle{{SIAM} J. Discret. Math.}
\bvolume{25}(\bissue{3}),
\bfpage{1443}--\blpage{1453}
(\byear{2011}).
\doiurl{10.1137/11082574}
\end{barticle}
\endbibitem

\bibitem{Guichar2004}
\begin{barticle}
\bauthor{\bsnm{Guichard}, \binits{D.R.}}:
\batitle{A lower bound for the domination number of complete grid graphs}.
\bjtitle{J. Combin. Math. Combin. Comput.}
\bvolume{49},
\bfpage{215}--\blpage{220}
(\byear{2004})
\end{barticle}
\endbibitem

\bibitem{Martinez2021}
\begin{barticle}
\bauthor{\bsnm{Mart\'inez}, \binits{J.A.}},
\bauthor{\bsnm{Garz\'on}, \binits{E.M.}},
\bauthor{\bsnm{Puertas}, \binits{M.L.}}:
\batitle{Powers of large matrices on \uppercase{GPU} platforms to compute the roman domination number of cylindrical graphs}.
\bjtitle{IEEE Access}
\bvolume{9},
\bfpage{29346}--\blpage{29355}
(\byear{2021}).
\doiurl{10.1109/ACCESS.2021.3058738}
\end{barticle}
\endbibitem

\bibitem{Pavlic2012}
\begin{barticle}
\bauthor{\bsnm{Pavli\v{c}}, \binits{P.}},
\bauthor{\bsnm{\v{Z}erovnik}, \binits{J.}}:
\batitle{Roman domination number of the cartesian products of paths and cycles}.
\bjtitle{Electron. J. Comb.}
\bvolume{19}(\bissue{3}),
\bfpage{19}
(\byear{2012})
\end{barticle}
\endbibitem

\bibitem{Kepner2011}
\begin{bbook}
\beditor{\bsnm{Kepner}, \binits{J.}},
\beditor{\bsnm{Gilbert}, \binits{J.R.}} (eds.):
\bbtitle{Graph Algorithms in the Language of Linear Algebra}.
\bsertitle{Software, environments, tools},
vol. \bseriesno{22}.
\bpublisher{{SIAM}},
\blocation{Philadelphia, USA}
(\byear{2011}).
\doiurl{10.1137/1.9780898719918}
\end{bbook}
\endbibitem

\bibitem{Dobo1990}
\begin{barticle}
\bauthor{\bsnm{Dobosiewicz}, \binits{W.}}:
\batitle{A more efficient algorithm for the min-plus multiplication}.
\bjtitle{Int. J. Comput. Math.}
\bvolume{32}(\bissue{1-2}),
\bfpage{49}--\blpage{60}
(\byear{1990}).
\doiurl{10.1080/00207169008803814}
\end{barticle}
\endbibitem

\bibitem{Felz2011}
\begin{barticle}
\bauthor{\bsnm{Felzenszwalb}, \binits{P.F.}},
\bauthor{\bsnm{McAuley}, \binits{J.J.}}:
\batitle{Fast inference with min-sum matrix product}.
\bjtitle{{IEEE} Trans. Pattern Anal. Mach. Intell.}
\bvolume{33}(\bissue{12}),
\bfpage{2549}--\blpage{2554}
(\byear{2011}).
\doiurl{10.1109/TPAMI.2011.121}
\end{barticle}
\endbibitem

\bibitem{Buluc2011}
\begin{barticle}
\bauthor{\bsnm{Buluç}, \binits{A.}},
\bauthor{\bsnm{Gilbert}, \binits{J.}}:
\batitle{The combinatorial \uppercase{BLAS}: Design, implementation, and applications}.
\bjtitle{The International Journal of High Performance Computing Applications}
\bvolume{25},
\bfpage{496}--\blpage{509}
(\byear{2011}).
\doiurl{10.1177/1094342011403516}
\end{barticle}
\endbibitem

\bibitem{Humayun2016}
\begin{botherref}
\oauthor{\bsnm{Humayun}, \binits{A.}},
\oauthor{\bsnm{Asif}, \binits{M.}},
\oauthor{\bsnm{Hanif}, \binits{M.K.}}:
{BTAS:} {A} library for tropical algebra.
CoRR
\textbf{abs/1701.04733}
(2017)
\end{botherref}
\endbibitem

\bibitem{Yang2020}
\begin{botherref}
\oauthor{\bsnm{Yang}, \binits{C.}},
\oauthor{\bsnm{Bulu{\c{c}}}, \binits{A.}},
\oauthor{\bsnm{Owens}, \binits{J.D.}}:
Graphblast: {A} high-performance linear algebra-based graph framework on the {GPU}.
CoRR
\textbf{abs/1908.01407}
(2019)
\end{botherref}
\endbibitem

\bibitem{Garzon2022}
\begin{botherref}
\oauthor{\bsnm{Garz{\'{o}}n}, \binits{E.M.}},
\oauthor{\bsnm{Mart{\'{\i}}nez}, \binits{J.A.}},
\oauthor{\bsnm{Moreno}, \binits{J.J.}},
\oauthor{\bsnm{Puertas}, \binits{M.L.}}:
On the 2-domination number of cylinders with small cycles.
Fund. Inform.
\textbf{accepted}
(2022)
\end{botherref}
\endbibitem

\bibitem{Carre1979}
\begin{bbook}
\bauthor{\bsnm{Carr\'e}, \binits{B.}}:
\bbtitle{Graphs and Networks}.
\bpublisher{Clarendon Press},
\blocation{Oxford, UK}
(\byear{1979})
\end{bbook}
\endbibitem

\bibitem{Spalding1998}
\begin{botherref}
\oauthor{\bsnm{Spalding}, \binits{A.}}:
Min-plus algebra and graph domination.
PhD thesis,
Dept. of Appl. Math., Univ. of Colorado,
Denver, CL, USA
(1998)
\end{botherref}
\endbibitem

\bibitem{OpenMP}
\begin{botherref}
The \uppercase{O}pen\uppercase{MP API} specification for parallel programming.
\url{https://www.openmp.org}.
Accessed: 2021-03-31
\end{botherref}
\endbibitem

\bibitem{nvidia2020}
\begin{botherref}
NVIDIA CUDA toolkit.
\url{https://developer.nvidia.com/cuda-math-library}.
Accessed: 2021-03-31
\end{botherref}
\endbibitem

\bibitem{nvidia2021}
\begin{botherref}
NVIDIA CUDA documentation.
\url{https://docs.nvidia.com/cuda/pdf/CUDA_C_Programming_Guide.pdf}.
Accessed: 2021-03-31
\end{botherref}
\endbibitem

\end{thebibliography}

\end{document}